# Mining Patterns in Networks using Homomorphism[*]

Anton Dries[†]    Siegfried Nijssen[‡]

**Abstract**

In recent years many algorithms have been developed for finding patterns in graphs and networks. A disadvantage of these algorithms is that they use subgraph isomorphism to determine the support of a graph pattern; subgraph isomorphism is a well-known NP complete problem. In this paper, we propose an alternative approach which mines tree patterns in networks by using subgraph *homomorphism*. The advantage of homomorphism is that it can be computed in polynomial time, which allows us to develop an algorithm that mines tree patterns in arbitrary graphs in incremental polynomial time. Homomorphism however entails two problems not found when using isomorphism: (1) two patterns of different size can be equivalent; (2) patterns of unbounded size can be frequent. In this paper we formalize these problems and study solutions that easily fit within our algorithm.

## 1  Introduction

Frequent pattern mining is arguably one of the most popular problems in the data mining literature. It involves enumerating all substructures satisfying a frequency constraint in data. Although initially studied on binary databases [1], such as market basket data, the task has been extended to many other types of data in recent years, including network and graph data [20, 13], motivated by the availability of large amounts of network data. Patterns were found to be particularly useful in such *structured data* as they can be used to build classifiers, track changes in network structure or cluster data [3, 2].

A key choice that has to be made in pattern mining algorithms is the definition of pattern support. Whether operating on sets of graphs [20, 16] or on networks [13, 2], the most common solutions in the literature are based on the use of *graph isomorphism*. We will illustrate this on the following example database.

---

[*]This work was partially supported by the Ministry of Science and Innovation of Spain (MICINN) under grant number TIN2009-14560-C03-01 and by a Postdoc and a project grant from the Research Foundation—Flanders, project "Principles of Patternset Mining".
[†]Universitat Pompeu Fabra, Barcelona, Spain
[‡]Katholieke Universiteit Leuven, Belgium

**Example 1.** *We assume given a database in which nodes correspond to authors, papers and keywords of papers. Directed edges indicate the authors and keywords of a paper.*

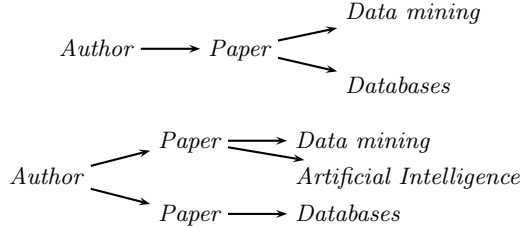

In the traditional transactional setting, each connected component of the graph above would be a different *transaction*; assuming a minimum support threshold of 2, the largest patterns found under subgraph isomorphism are:

- Author → Paper → Data mining
- Author → Paper → Databases

Both graphs are subgraph isomorphic to two transactions in the data and are hence frequent. When considering the graph as one large network, support measures based on overlap can be used [13]. Also in this case, however, these would be the largest patterns that can be found.

One pattern that is *not* found using subgraph isomorphism is hence the following:

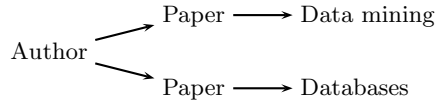

The reason is that the two Paper nodes in the pattern may not be mapped to the same node in the data at the same time, i.e. an isomorphism is an *injective* mapping. Consequently, the pattern that both authors have a paper in data mining and in databases can not be found when using subgraph isomorphism.

The technique that we propose in this paper addresses this specific injectivity requirement in existing graph mining algorithms. The key idea is to develop a pattern mining algorithm in which support is based on subgraph *homomorphism*, in which the mapping from a pattern to the data no longer needs to be injective.

Graph *homomorphism* is a well-studied topic in graph theory [7]. It is closely related to the way queries are usually evaluated in relational databases and to the $\theta$-subsumption procedure commonly used in inductive logic programming [5]. Indeed, the common use of homomorphisms in many areas in computer science can be seen as evidence that the use of homomorphism is intuitive to many people.

Another reason for the popularity of homomorphism is the fact that homomorphisms can be calculated more efficiently than isomorphisms. For instance, homomorphisms of a rooted tree in a graph can be calculated in polynomial time, while the corresponding problem for isomorphisms is NP complete [7]. In particular when dealing with large data, this is a crucial advantage well-recognized by the database community [6].

Nevertheless, in pattern mining the setting has rarely been studied, despite its potential computational advantages. There may be several reasons for this, which we will discuss in more detail in Section 2. Essentially, most of these problems are of a technical nature:

- trees of a different size can be homomorphic with each other, making the development of level-wise algorithms difficult;

- infinite numbers of patterns can be frequent when using homomorphisms on cyclic data without further constraints;

Indeed, the few existing pattern mining algorithms that support the graph homomorphism setting, do not find all frequent patterns even if this number is finite [5], cannot be applied on networks, or are not known to achieve incremental polynomial time enumeration even if restricted to rooted trees [8, 9].

In this paper, we develop a new algorithm that overcomes these problems. We will prove that this algorithm is able to find all frequent rooted trees under homomorphism and that it finds them in incremental polynomial time. Novel contributions of this paper are hence:

- we present the first algorithm that finds frequent rooted trees under homomorphism in incremental polynomial time, both in a transactional and in a network setting;

- we present novel measures of support, allowing the discovery of patterns in cyclic network data without transactions;

- we present the first algorithm for finding closed and maximal patterns under these settings;

- we present an extension of this algorithm which allows for pattern mining under syntactical constraints.

Syntactical constraints will turn out to be useful in an application of our algorithm on bibliographic data.

The remainder of the paper is organized as follows. In Section 2 we will introduce the basic graph principles that we need in this paper and provide more details on the technical challenges of pattern mining under homomorphism. In Section 3 we will introduce our algorithm for enumerating trees under homomorphism, as well as the basic algorithm for evaluating support. In Section 4 we introduce constraints that aim at avoiding unbounded patterns. Section 5 discusses the extension towards closed and maximal pattern mining. Section 7 discusses important related work in detail. Section 8 provides experiments while Section 9 concludes.

## 2 Preliminaries

### 2.1 Graphs, Trees and Homomorphisms

A *labeled directed* graph $G = (V_G, E_G, \lambda_G)$ (or *graph* for short in this paper) consists of a set of nodes $V_G$, a set of edges $E_G \subseteq V_G \times V_G$ and a labeling function $\lambda_G : V_G \cup E_G \to \Sigma$ that maps each node or edge of the graph to an element of the alphabet $\Sigma$. By $\mathcal{G}_\Sigma^D$ we denote the set of all graphs over the alphabet $\Sigma$. In case the graph is unlabeled, we drop $\lambda_G$.

A graph $G_1$ is subgraph *homomorphic* with another graph $G_2$, denoted by $G_2 \succeq G_1$, if there exists a function $\varphi$ from nodes in $V_{G_1}$ to nodes in $V_{G_2}$, such that:

- $\forall v \in V_{G_1} : \lambda_{G_1}(v) = \lambda_{G_2}(\varphi(v))$;
- $\forall (v_1, v_2) \in E_{G_1} : (\varphi(v_1), \varphi(v_2)) \in E_{G_2}$;
- $\forall (v_1, v_2) \in E_{G_1} : \lambda_{G_1}((v_1, v_2)) = \lambda_{G_2}((\varphi(v_1), \varphi(v_2)))$.

We also denote this as $G_2 \succeq_\varphi G_1$, for a given $\varphi$ that is a homomorphism. A graph $G_1$ is subgraph *isomorphic* with another graph $G_2$, denoted by $G_2 \succeq_\varphi^I G_1$, if $G_2 \succeq_\varphi G_1$, and $\varphi$ is *injective*, i.e., if $v_1 \neq v_2$, then $\varphi(v_1) \neq \varphi(v_2)$.

Given two graphs $G_1$ and $G_2$, if $G_1 \succeq G_2$ and $G_2 \succeq G_1$, we denote this by $G_1 \equiv G_2$; similarly we also define $G_1 \equiv^I G_2$. The equivalence class under homomorphism of a graph is defined as follows:

$$[G] = \{G' \in \mathcal{G}_\Sigma^D \mid G' \equiv G\}.$$

A labeled *rooted unordered tree* $T = (V_T, E_T, \lambda_T)$ is a graph which fulfills the additional requirement that there is exactly one node $v \in V_T$ such that:

- there is no edge $(v', v) \in E_T$;
- for all other nodes $v' \in V_T$ there is exactly one node $v'' \in V_T$ such that $(v'', v') \in E_T$.

Node $v$ is called the root of the tree; node $v''$ is the parent of node $v'$ while node $v'$ is a child of node $v''$. Subgraph isomorphism and homomorphism carry over from graphs.

Without loss of generality, we will assume from now on that node $v_0^T$ is the root of a tree. A *root preserving subtree homomorphism* of tree $T_1$ into $T_2$ is a subtree homomorphism in which $\varphi(v_0^{T_1}) = v_0^{T_2}$; it is denoted by $T_2 \succeq^R T_1$. Similarly, a root preserving isomorphism is denoted by $T_2 \succeq^{R,I} T_1$. We say that $T_2$ is a *specialization* of $T_1$, while $T_1$ is a *generalization* of $T_2$, if $T_2 \succeq^R T_1$.

For a given (infinite) set of nodes $\mathcal{V} = \{v_0, v_1, \ldots\}$ and a finite set of labels $\Sigma$ we denote by $\mathcal{G}_{\mathcal{V},\Sigma}^T$ the set of all (finite) rooted unordered trees over these nodes and labels.

## 2.2 Differences between Homomorphism and Isomorphism

Graph isomorphism and homomorphism behave differently. These differences lead to two important problems.

**Problem 1.** *An infinite number of graphs, each pair of which is not isomorphic, can all be homomorphic with each other.*

Consider for instance this infinite set of directed graphs:

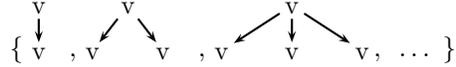

Each trees in this set is homomorphic to all other trees and hence conveys the same information. Finding one of these trees is sufficient. The consequence is that we cannot straightforwardly reuse the enumeration process of an isomorphism-based graph mining algorithm: an existing graph mining algorithm would try to enumerate all these patterns and produces redundant results.

**Problem 2.** *An infinite number of graphs, each pair of which is not homomorphic with each other, can be subgraph homomorphic with a finite graph.*

As an example consider this infinite set of graphs:

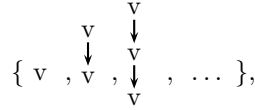

then each of these graphs is subgraph homomorphic with this graph:

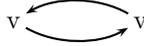

The cause of this problem is the presence of a cycle in the data. If the data is not cyclic, this problem does not occur.

These problems motivate our problem definition in the next section.

## 2.3 High-Level Problem Description

In this section we will introduce the basic problem of frequent rooted tree mining under homomorphism; in a later section we will define extended problem settings, such as closed and maximal pattern mining. Let us define the *support* of a tree $T$ in a graph $G$ as follows.

**Definition 1** (Image)**.** *Given a tree $T$ and a graph $G$, the image of a node $v \in V_T$ is:*
$$img_{T \to G}(v) = \{v' \in V_G \mid \exists \varphi : \varphi(v) = v', G \succeq_\varphi T\}$$

**Definition 2** (Support)**.** *The support of a tree can now be defined as the size of the image set of the root:*
$$freq(T, G) = \left|img_{T \to G}(v_0^T)\right|.$$

*Using this support we can define a support constraint:*

$$\varphi(T, G) = (\mathit{freq}(T, G) \geq \theta)$$

*for some threshold $\theta$.*

Note that this definition assumes that the data is one large network. It can easily be modified to a transactional setting.

Intuitively one could now be interested in finding the set of *frequent* trees

$$\mathcal{F} = \{T \in \mathcal{G}_\Sigma^T \mid \varphi(T, G) \text{ is true }\}.$$

Note that this support definition is *anti-monotonic* under root-preserving subtree homomorphism, i.e. if $T_1 \succeq^R T_2$ and $\varphi(T_1, G)$ is true, then also $\varphi(T_2, G)$ is true. Furthermore, any pattern that would be frequent when using isomorphism would also be frequent when using homomorphism.

Due to Problem 1 the set $\mathcal{F}$ is infinite. Hence, we need to refine this problem statement. Essentially, we wish to find only one *representative* for each equivalence class of trees. Let us assume we have a function $r(T)$ that computes one representative for a set of equivalent trees $[T]$, i.e. for all $T' \in [T]$: $r(T') = r(T)$. Then we define a reduced space of trees

$$r(\mathcal{G}_\Sigma^T) = \{r(T) \mid T \in \mathcal{G}_\Sigma^T\}$$

and look for subtrees in

$$\mathcal{F} = \{T \in r(\mathcal{G}_\Sigma^T) \mid \mathit{freq}(T, G) \geq \theta\}.$$

In Section 3 we propose to use *core* trees as representatives, and we show how to enumerate such core trees.

Due to Problem 1 even the reduced space of trees can contain an infinite number of frequent patterns. To avoid this problem, we propose to impose additional constraints that patterns need to fulfill in order to be part of the result set. These constraints have to fulfill the following property and are introduced in detail in Section 4.

**Definition 3** (Finiteness)**.** *An anti-monotonic constraint $\varphi$ ensures finiteness iff for any graph $G$ there is a maximum on the size of the tree $T$ for which $\varphi(T, G)$ is true.*

We are initially only interested in finding core patterns under an anti-monotonic constraint that ensures finiteness.

## 3 Enumerating Patterns

### 3.1 Refined Problem Definition

Essentially, we propose to enumerate core trees as representative patterns. A tree is a *core* if no subtree can be identified that is equivalent with it.

**Definition 4** (Core tree). *A tree $T$ is a core iff there is no tree $T'$ with: (1) $T \succeq^I T'$; (2) $T \not\equiv^I T'$; (3) $T \equiv T'$.*

Of the below two trees, tree $T_1$ is a core; tree $T_2$ is not a core, as it contains the equivalent tree $T_1$ (labels are indicated in black, node identifiers in gray).

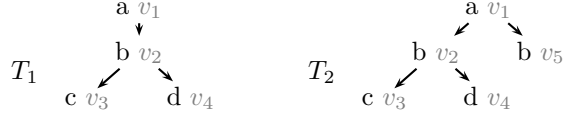

We will restrict our enumeration to core trees as much as possible. If a tree is not core, we will say that it is *reducible* to a core tree.

For any given tree $T = (V, E, \lambda)$ and given node $v \in V$, we will denote by $subtree(v, T)$ the complete subtree below $v$ in $T$, i.e. the tree consisting of all nodes that are a descendant of $v$ in $T$ and all edges connecting these nodes, as well as the path leading from the root of $T$ to node $v$.

In our example, $subtree(v_2, T_2) = T_1$.

An interesting feature of core trees is the following.

**Lemma 1.** *A tree $T$ is a core iff for every pair of siblings $v_1$ and $v_2$ in $T$, neither $subtree(v_1, T) \succeq^I subtree(v_2, T)$ nor $subtree(v_2, T) \succeq^I subtree(v_1, T)$. If two trees are a core $T_1 \equiv T_2$ implies $T_1 \equiv^I T_2$.*

*Proof.* First we prove that if a tree is a core, then for every pair of siblings $v_1$ and $v_2$ in $T$, neither $subtree(v_1, T) \succeq^I subtree(v_2, T)$ nor $subtree(v_2, T) \succeq^I subtree(v_1, T)$.

Assume that we would have $subtree(v_1, T) \succeq^I subtree(v_2, T)$. Then consider the tree $T'$ without $subtree(v_2, T)$ (except for the path to node $v_2$). Clearly we would have $T \succeq^R T'$; however also $T' \succeq^R T$ is the case, as the nodes in $subtree(v_2, T)$ in $T$ could be mapped to the nodes in $subtree(v_1, T')$. As we have identified a tree for which $T \succeq^R T'$, the assumption that $subtree(v_1, T) \succeq^I subtree(v_2, T)$ was incorrect.

The other way around, if the tree is not a core, we can assume equivalence with a root-preserving subgraph isomorphic subtree $T'$. The only possibility for such equivalence with a smaller tree is that we map at least two sibling nodes of $T$ to the same node $v'$ in tree $T'$. As the subtree $subtree(v', T')$ is also a subtree of $T$, this must mean that the subtree $subtree(v_1, T)$ for one node $v_1$ of $T$ is mapped into a subtree $subtree(v_2, T)$ for some node $v_2 \in V_T$. □

Note that as we only enumerate core trees, certain tree structures will not be found under homomorphism-based mining that could be found using isomorphism-based algorithms. According to this lemma, the patterns that will not be found in our algorithm will always contain a subtree that is subgraph isomorphic to another subtree within the pattern. Indeed, when using subgraph isomorphism in practice [2] one often finds patterns that indirectly encode degree constraints. Such patterns are avoided by using homomorphisms.

When enumerating trees in today's sequential computers, we need to develop a representation of trees in which the nodes are ordered. The second problem that we need to deal with is that different *ordered trees* can represent the same unordered tree. Here we reuse ideas from earlier tree mining algorithms [16, 4].

For simplicity, we assume that a tree is *ordered* if the nodes are numbered as $v_1, \ldots, v_n$ in a depth-first order. The following are two ordered trees:

$$T_1 = (\{v_1, v_2, v_3, v_4\}, \{(v_1, v_2), (v_2, v_3), (v_1, v_4)\})$$

and

$$T_2 = (\{v_1, v_2, v_3, v_4\}, \{(v_1, v_2), (v_1, v_3), (v_3, v_4)\}).$$

These trees are equivalent to each other under isomorphism, but represent different orders in which the children of $v_1$ are enumerated. The *right-most path* of an ordered tree is the path from the root to the last node in the tree.

To ensure that only one of these two orders is enumerated, we will use a *canonical code*. Canonical codes are common in graph and tree mining; here we will use the canonical code of [16], which works as follows.

The *code* of an *ordered tree* is computed by traversing the tree in depth-first order, and printing for every node its depth and its label. For our example trees we obtain the following codes:

$$\ell(T_1) = 0a1a2a1a \text{ and } \ell(T_2) = 0a1a1a2a.$$

Among all codes that can be obtained for a tree, the *canonical code* is the one which is lexicographically the highest[1]. In our example, as $\ell(T_1) \geq \ell(T_2)$, where $\geq$ denotes lexicographical comparison, $\ell(T_1)$ is canonical; $\ell(T_2)$ is not. A characteristic of this canonical code is that trees look 'left-heavy': if there were no labels, the longest path would always be the left-most path in the tree.

The refined enumeration problem is now the following.

**Definition 5** (Enumeration under Homomorphism). *Given a constraint $\varphi$ that ensures finiteness, find the set:*

$$\{T \in \mathcal{G}_\Sigma^T \mid \varphi(T, G) \text{ is true, } T \text{ is canonical and a core}\}.$$

### 3.2 Algorithm

In our algorithm we exploit several properties of the above canonical code. We repeat these properties here; proofs for the first three claims can be found in [14, 15].

**Property 1.** *Every prefix of a canonical code representation of a tree also is the canonical representation of the corresponding tree.*

For instance, the prefixes $0a$, $0a1a$, $0a1a2a$ of $\ell(T_1) = 0a1a2a1a$ are also canonical codes.

**Property 2.** *A tree $T$ is canonical iff for any pair of siblings $v_i$ and $v_j$ ($i < j$): $\ell(subtree(v_i, T)) \geq \ell(subtree(v_j, T))$.*

---

[1] Here some order on the labels is assumed

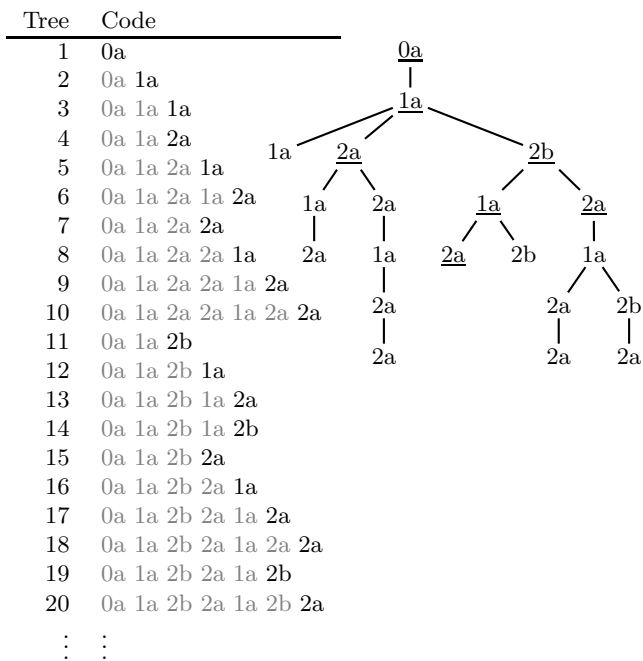

| Tree | Code |
|---|---|
| 1 | 0a |
| 2 | 0a 1a |
| 3 | 0a 1a 1a |
| 4 | 0a 1a 2a |
| 5 | 0a 1a 2a 1a |
| 6 | 0a 1a 2a 1a 2a |
| 7 | 0a 1a 2a 2a |
| 8 | 0a 1a 2a 2a 1a |
| 9 | 0a 1a 2a 2a 1a 2a |
| 10 | 0a 1a 2a 2a 1a 2a 2a |
| 11 | 0a 1a 2b |
| 12 | 0a 1a 2b 1a |
| 13 | 0a 1a 2b 1a 2a |
| 14 | 0a 1a 2b 1a 2b |
| 15 | 0a 1a 2b 2a |
| 16 | 0a 1a 2b 2a 1a |
| 17 | 0a 1a 2b 2a 1a 2a |
| 18 | 0a 1a 2b 2a 1a 2a 2a |
| 19 | 0a 1a 2b 2a 1a 2b |
| 20 | 0a 1a 2b 2a 1a 2b 2a |
| ⋮ | ⋮ |

Figure 1: The first 20 canonical codes for trees upto depth 2, degree 2, and labels {a,b} (left); a visualization of the trie data structure storing these trees, not taking into account homomorphism (right); nodes that are stored also under homomorphism are underlined.

**Property 3.** *If $T_1 \succeq^{R,I} T_2$, then $\ell(T_1) \geq \ell(T_2)$.*

This property is especially important, as it states that for any tree $T_1$ all subtrees must have been enumerated before this tree, including the core trees when using homomorphism, if we enumerate the patterns in the order of their codes.

We enumerate trees depth-first, from smallest to largest according to the lexicographical order of canonical codes, by using Algorithm 1 based on the algorithm of [15]. This algorithm was developed for enumerating trees under an isomorphism relationship. Ignoring the problem of homomorphisms for the moment, the first trees enumerated up to depth 2, degree 2 and for labels {a,b} are given in Figure 1. Trees are grown by extending the codes of trees, where the essential observation is that in a canonical code we can only add edges to the rightmost path of a tree.

To calculate whether a given new code is canonical [14, 15] proposed an $O(1)$ algorithm which we do not elaborate on here as we will not use it later on.

An important observation when applying this algorithm in a homomorphism context is that it is sometimes necessary to enumerate trees which are reducible in order to reach a tree which is a core. An example is tree number 12, which is not a core, but needed to reach tree 13 in Figure 1.

**Algorithm 1** Tree enumeration(code)
---
**if** *code* is not canonical or does not satisfy constraint $\varphi$ **then**
  return
output *code*
let $d\sigma$ be the last (depth,label) pair in *code*
**for** $i := 1$ to $d$ **do**
  **for** $\sigma \in \Sigma$ in increasing order **do**
    Tree enumeration(*code* $\cdot i \cdot \sigma$)
---

The issue that we need to address is how to avoid as much as possible that Algorithm 1 enumerates trees that are reducible to a core tree, but that are considered in the hope of reaching a tree which is a core and satisfies the constraints later on. In the following we will show such a method for which we can prove a good theoretical performance.

The first idea is to exploit the fact that certain trees can never be extended into a core tree.

**Property 4** (Unavoidably Reducible). *Given a canonical tree $T$, if we can identify two siblings $v_1$ and $v_2$ such that $subtree(v_1, T) \succ^R subtree(v_2, T)$, while both $v_1$ and $v_2$ are not on the right-most path of the tree, then no canonical code $\ell(T')$ which has $\ell(T)$ as a prefix is a core.*

*Proof.* This follows from the fact that we grow trees by connecting nodes to the right-most path and from Lemma 1. □

The second idea is that we store all tree-patterns that are found in the enumeration in a trie data structure (see Figure 1); each node in this data structure corresponds to a tree pattern represented by its canonical code. The children of a node in the data structure essentially represent those trees that are canonical children in the enumeration tree.

When dealing with homomorphism, we will maintain this data structure such that any given node is only stored if one of its descendants is a core tree. In our example, tree number 7 is for instance not stored, as none of its descendants is a core tree.

We use this data structure to guide the traversal of the search space. While growing trees in Algorithm 1, we traverse the already constructed part of the trie to eliminate extensions that are not known to lead to trees that are frequent or a core; essentially, hence, we here exploit the fact that frequency is an anti-monotonic property.

We perform this traversal of the trie as follows. The node on the *right-most* path of the pattern with the lowest level and more than one child will be referred to as the *split node*. If we now remove *the first subtree* below this node, we obtain an ordered tree which is also a canonical tree (by application

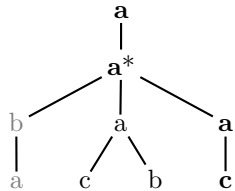

0a 1a 2b 3a 2a 3c 3b 2a 3c

Figure 2: Example of a canonical tree pattern, its rightmost path (bold) and its split node (*)

of Property 2), and which —if it is satisfies the anti-monotonic constraint— we must have entered in the trie earlier according to Property 3.

This is illustrated in Figure 2. If we remove the nodes which are marked in gray, we obtain a code which is also canonical. We can search for this code in the data structure; only the children of this code in the data structure are promising extensions of the current code.

A special case is a pattern in which we cannot identify a split node, as it is consists of one path. In this case we treat every node on the path iteratively as a split node, starting from the shallowest node.

Note that due to the order in which we are enumerating trees, and the fact that we only allow extensions which are part of the data structure, we do not need to perform explicit checks whether a code is canonical.

This leads us to the algorithm given in Algorithm 2. This algorithm is called in principle for each $\sigma \in \Sigma$, where $\Sigma$ is the set of labels occurring in the data, in increasing order.

In this algorithm we need to calculate whether a tree is a core, and whether reducibility is avoidable. In principle, we could determine this for each pattern tree from scratch using a polynomial algorithm. However, by reusing information we can do this more efficiently; we calculate incrementally whether a pattern tree is a core and whether a pattern tree is unavoidably reducible.

First, we observe that a path-like tree is always a core; we only need to consider trees which have a split node. If we remove the left-most subtree of the split node, we obtain a tree which was stored earlier in the trie data structure, and for which we have already stored whether it is a core; if this earlier tree was not a core, the current tree, which includes this tree, is not a core either. If the earlier stored tree was a core, reducibility can only be caused by the presence of the left-most subtree, where there are two possible reasons:

- the left-most subtree is not a core itself;
- one of the other subtrees of the split-node is a subtree of the left-most subtree (Lemma 1).

**Algorithm 2** Tree enumeration(*code*)
---
**if** *code* does not satisfy constraint $\varphi$ **or**
    *code* is unavoidably reducible **then**
  return
**if** *code* is a core **then**
  insert *code* into trie data structure
  output pattern
**if** *code* represents a path **then**
  *split nodes* ← { all nodes on the path }
**else**
  *split nodes* ← { node on right-most path
    with lowest depth and more than one child }
**for all** *split nodes* in increasing depth **do**
  let *code′* be the code of the tree obtained when removing
    the left-most subtree of *split node*
  **for all** children ($d\sigma$) of *code′* in trie data structure **do**
    Tree enumeration(*code* $\cdot d \cdot \sigma$)
**if** *code* represents a path **then**
  let $d$ be the depth of the deepest node $+1$
  **for** $\sigma \in \Sigma$ in increasing order **do**
    Tree enumeration(*code* $\cdot d \cdot \sigma$)

---

We can also exclude many of these possibilities, as we ensure during the enumeration process that only trees with avoidable reducibility are considered (note that we check reducibility only *after* we have checked unavoidable reducibility). If any subtree which is not the right-most subtree of the split node could be mapped to the left-most subtree, or the left-most subtree would not be a core, the tree would be unavoidably reducible, and the tree would have been pruned earlier.

Hence, we only need to check whether the right-most subtree of the split node can be mapped to the left-most subtree. To calculate this efficiently we use an incremental algorithm similar to the algorithm used to evaluate the support of a tree pattern. This algorithm is discussed in Section 3.4.

Similarly, we can also calculate incrementally whether reducibility is avoidable. First, we observe that we can assume that the parent pattern in the enumeration is already avoidably reducible. Whether this is still the case for the current pattern is easily determined:

- if the new node was inserted above the split node of the parent tree, or was connected to the split node, and this parent tree was not a core, this creates an unavoidable reducibility;

- if the new node was inserted below the split node, any unavoidable reducibility would have to be below this split node. The subtree below this split node corresponds to a tree that has been stored earlier, however, and we can assume that no tree with unavoidable reducibility is stored.

### 3.3 Complexity Analysis

We provide an analysis of the complexity of the above algorithm here, assuming that both the support constraint and the reducibility of a pattern can be evaluated in polynomial time (we will see in the next section that this is the case).

Given that the potential number of patterns is exponential, our analysis is based on the amount of computation needed per pattern enumerated. We show that the algorithm achieves *incremental polynomial time* complexity.

**Definition 6** (Incremental Polynomial Time [12]). *An algorithm to enumerate items $a_1, a_2, \ldots, a_p$ runs in* incremental polynomial time *if*

- *it iterates the following procedure for $i = 1, 2, \ldots, p$: output the ith item $a_i$ from the knowledge of its input and items $a_1, a_2, \ldots, a_{i-1}$ generated so far;*

- *the time required for the ith iteration is polynomial in the input length and the sizes of $a_1$, $a_2$, $\ldots$, $a_{i-1}$.*

The main argument is that our algorithm only considers extensions of a path that can be obtained by copying earlier trees below the path. Hence, to reach a next pattern, a number of possibilities bounded by the number of trees stored is considered, which in its turn is polynomial in the number of patterns listed earlier.

More formally, given a certain (core) pattern tree, the next (core) pattern tree satisfying the given constraint $\varphi(T, G)$ can be of two kinds. (1) The next pattern tree is a descendant of the current pattern in the enumeration tree. Descendants can be reached in two ways: by adding trees found earlier (and stored in the trie) to the right-most path, or by adding nodes below the left-most node in case the tree only has one path. In the first case the number of trees that can be added to the right-most path is bounded by the number of trees found earlier; in the second case, the number of labels we can add below the path is bounded by the number of labels in the data. Overall, the number of attempted refinements before a frequent refinement is reached is hence bounded polynomially. (2) The next pattern tree is a descendant of an ancestor of the current pattern. The number of such ancestors is bounded by the number of nodes in the pattern. For each such ancestor, we can apply the argument above.

Please note that a more desirable time complexity would be polynomial *delay*, in which case the time needed to enumerate the next pattern would be bounded by a polynomial that is independent of the number of patterns enumerated till that moment. It is still an open question whether enumeration with this type of complexity can be achieved.

### 3.4 Evaluating Support

Efficiently evaluating support is critical for pattern mining algorithms; indeed an important motivation for our work is that we can evaluate support efficiently for the case of homomorphism. In this section we show how we can evaluate the support of a pattern.

In principle we could evaluate the support of a rooted tree from scratch by means of known algorithms for graph homomorphism [6], or by means of SQL database queries [9]. However, a more efficient approach is obtained by evaluating support incrementally. The main idea is that for every node $v$ on the right-most path, we maintain its image $img_{T \to G}(v)$ in the data. If we extend a pattern by adding a node to the right-most path (yielding a pattern $T'$), the image of each node on the right-most path of $T'$ is obtained as follows:

- the image for the new node $v_{new}$ with label $\sigma$ is created by traversing the image of its parent $v_{parent}$; for each $v' \in img_{T \to G}(v_{parent})$ those nodes $v''$ with $(v', v'') \in E_G$ and $\lambda_G(v'') = \sigma$ are inserted into $img_{T' \to G}(v_{new})$.

- the image of the other nodes is update sequentially in a bottom-up fashion, starting from the node $v_{parent}$. Essentially, any node in $img_{T \to G}(v_{parent})$ which is not connected (in $G$) to at least one node in $img_{T' \to G}(v_{new})$ is removed.

The size of the image of the root, $|img_{T' \to G}(v_0)|$, corresponds to the support of the new pattern.

Note that each image list is no larger than the number of nodes in the data; updating each list proceeds in polynomial time. Overall, this provides us with polynomial time support evaluation.

## 4 Constraints for Cyclic Data

The above algorithm can safely be applied to graphs which are directed and acyclic; in this case our algorithm is the first algorithm for enumerating rooted trees in incremental polynomial time. However, when applied to cyclic graphs, such as most information networks, we are faced with Problem 2. In this section we will introduce a general class of *path-based* constraints that can be used in addition to support constraints to avoid this problem, while being easy to evaluate in our algorithm.

**Definition 7.** *A constraint $\varphi(T, G)$ is* path-based *if it can be expressed as a constraint over all root-paths in the tree:*

$$\varphi(T, G) = \bigwedge_{v \in V_T} \varphi'(path(v, T), G),$$

*where $path(v, T)$ denotes the tree which consists only of the path leading from the root of $T$ to the node $v$, and $\varphi'$ is a* path constraint *that can only be evaluated on path patterns.*

If a path constraint ensures finiteness of paths (see Definition 3), the above path-based constraint ensures finiteness of trees.

Note that a path-based constraint is anti-monotonic: any supertree of a tree which does not satisfy a path-based constraint will not satisfy it either.

A path-based constraint is easily integrated in our algorithm: essentially, when we are growing paths, we need to check whether the paths satisfy the path constraint. Other paths are only added in a tree if they have been enumerated earlier, and hence have been found to satisfy the path constraint earlier. Hence, as soon as a pattern has a split node, checking the path-based constraint is no longer needed.

Below we provide an overview of possible path constraints on which tree constraints can be based.

**Depth Constraint** The depth constraint is defined by:

$$\varphi'(P, G) \Leftrightarrow n \leq \delta,$$

where $n$ is the number of nodes in path $P$.

**Additional Cover Constraint** This constraint determines if the last node of the path adds information with respect to the union of earlier nodes in the path:

$$\varphi'(P, G) \Leftrightarrow \left( img_{P \to G}(v_n) \nsubseteq \bigcup_{0 \leq i < n} img_{P \to G}(v_i) \right),$$

where $V_P = \{v_0, \ldots, v_n\}$.

Intuitively, this constraint avoids that many nodes of a pattern are mapped to the same set of nodes in the data. This is the default constraint that we will be using in our experiments.

**Label Constraint** The label constraint states that no two nodes on a path may have the same label:

$$\forall v, v' \in V_P : v \neq v' \implies \lambda_P(v) \neq \lambda_P(v').$$

Note that this constraint is sufficient to find the patterns in the introduction (Section 1). Even if this constraint is applied, subgraph isomorphism remains NP-complete, and hence the use of homomorphism is beneficial. We can prove this by showing that we can reduce the NP-complete independent set problem to a problem of subgraph isomorphism under the label constraint.

**Definition 8.** *(Independent Sets) Given a graph $G = (V, E)$ an independent set is a subset of nodes $V' \subseteq V$ such that for no $v_1, v_2 \in V'$: $(v_1, v_2) \in E$. The problem of finding a k-independent set is the problem of finding an independent set of size $k$.*

The independent set problem is a well-known NP complete problem.

**Theorem 1.** *Subgraph isomorphism under the label constraint is NP complete.*

*Proof.* We can encode the independent set problem in terms of the subgraph isomorphism problem as follows. The data graph is a directed acyclic graph $D$ of three levels. The levels are the following:

- level 1 in $D$ contains one node (the root), having label "R";

- level 2 in $D$ contains one node for each node in the graph $G$, each node being connected to the root; these nodes are labeled with label "N"

- level 3 in $D$ contains nodes labeled with label "N2", in the following configuration:

  - one node for each node in the graph $G$; each such node is connected to all nodes in level 2 to which it is connected in $G$, as well as the node itself; the idea is that these are the nodes that would be covered if the node in the original node is selected, which includes all neighbors and the node itself;
  
  - let $m$ be the maximum degree of the nodes in $G$. For each node with degree $d < m$, $m - d$ additional nodes are added in level 3, which are only connected to the corresponding node in level 2.

The query graph consists of three levels as well:

- level 1 with one root with label "R";

- level 2 with $k$ nodes labeled with label "N", each connected to the root;

- level 3 with $km$ nodes labeled with label "N2", such that each node in level 2 has $m$ children.

This transformation is clearly polynomial; furthermore, on each path in the query tree all labels are unique.

If we find a subgraph isomorphism for this query graph in the data graph, we have selected $k$ nodes in level 2 of the data graph, such that no two nodes in level 3 of the query graph are mapped to the same node in the data graph; this is only possible if we selected nodes that do not share an edge, and hence we have solved the independent set problem. □

## 5 Maximality

As most other frequent pattern mining algorithms, the above algorithm may generate many frequent patterns. One means to reduce the number of patterns is by restricting the output to patterns that are either *maximal* or *closed*. A *maximal frequent pattern* is a pattern for which no specialization satisfies the given constraints; a *closed frequent pattern* is a pattern for which no specialization has the same support.

We can filter the output of the above algorithm to obtain such patterns. Compared to tree mining algorithms in the literature [4], a complicating issue is that we can also specialize by *merging* nodes in a tree:

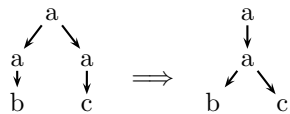

Note that merging does not affect the set of paths present in the tree, and hence can never influence the evaluation of a path-based constraint.

We now discuss how we find maximal patterns by post-processing the set of patterns found above; subsequently, we discuss how to make the search more efficient by using additional pruning criteria.

## 5.1 Post-Processing

To obtain a set of maximal patterns in a naïve way, we do the following in post-processing:

- for every pattern, we determine which generalizations can be obtained by removing a leaf; we mark such generalizations as not maximal;

- for every pattern, we determine which specializations can be obtained by considering a *merge* of every pair of siblings. Merging proceeds as follows: (1) two siblings are replaced by one node, below which we copy the subtrees of the two original trees; (2) we reduce a tree to its core to remove redundant subtrees; (3) we compute its canonical form; (4) if the resulting code is a frequent pattern, the pattern is marked as non-maximal.

The reason for merging is that generating specializations by *splitting* a tree is a more costly procedure, as there are many possible ways to split a certain node in a tree.

## 5.2 Pruning during the Search

Several optimizations can be applied to avoid the enumeration of patterns during the search that cannot be maximal.

The main idea is illustrated by the following example. Assume we are enumerating the following tree $T$, where we have added superscripts to give different names to nodes labeled with a:

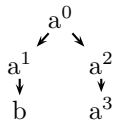

Then we can prune this tree in the following cases.

**Test 1** Let $T'$ be the last ancestor in the enumeration of $T$ in which node $a^1$ was on the right-most path, i.e. the trees with $a^2$ and $a^3$ missing. Then we can assume that we still know the image of $a^1$ of this earlier pattern due to the depth-first enumeration. We prune the pattern $T$ if:

$$img_{T \to G}(a^2) \subseteq img_{T' \to G}(a^1)$$

The reason is that any subtree that is inserted below node $a^2$ (yielding a tree $T''$), can additionally be attached below node $a^1$; any such pattern will be more specific and can not have a support lower than $T''$.

**Test 2** Similarly, we can also prune if

$$img_{T \to G}(a^2) \supseteq img_{T' \to G}(a^1);$$

however in this case an additional condition is required; after all, adding nodes below $a^2$ may remove some of the images of $a^2$, yielding situations in which $a^2$'s image is no longer a superset of $a^1$'s. The solution to this is to apply such pruning only once a new node is connected to $a^0$; due to our enumeration process we know at this point that the tree below $a^2$ will not grow any more.

**In general** We can describe these tests formally as follows. Assume given a pattern tree $T$. Let $V = \{v_0, \ldots, v_n\}$ be its set of nodes, in the order of the canonical form. Furthermore, let $R_T(i)$ be the index of node at depth $i$ on the rightmost path of $T$. Let $T[i]$ denote the tree which contains only nodes $\{v_0, \ldots, v_i\}$ of tree $T$, i.e. $T[i]$ is an ancestor of $T$ in the search space, which is also canonical due to our definition of the canonical form.

Then, according to *test 1*, we prune a pattern $T$ if there are nodes $u, u' \in V_T$ for which:

- $u = v_{R_T(i)}$ for a certain $i$, i.e. $u$ is a node on the rightmost path;

- $u'$ is the left-hand sibling of $u$, if there is any such node;

- $\lambda_T(u) = \lambda_T(u')$, i.e. $u$ and $u'$ have the same label;

- $img_{T \to G}(u) \subseteq img_{T[R_T(i)-1] \to G}(u')$; here $T[R_T(i) - 1]$ is the last ancestor of $T$ in which $u'$ was on the rightmost path.

The main argument for this test is that if we now consider any tree $T'$ which contains additional nodes below node $u$ in the pattern, we will still have the following property:

$$img_{T' \to G}(u) \subseteq img_{T' \to G}(u').$$

The reason for this is that the only cause for losing a node $x$ in $img_{T' \to G}(u')$ is that we lost a node in the image of the parent of $u'$. However, then we must have also lost the node $x$ in $img_{T' \to G}(u)$, as $u$ and $u'$ have the same parent, and hence the subset relation still holds.

As $img_{T'\to G}(u) \subseteq img_{T'\to G}(u')$ for every tree $T'$, we can put the subtrees below node $u$ in tree pattern $T'$ always below node $u'$; the resulting pattern will also be frequent, while being more specific.

Similarly, according to *test 2*, we prune a pattern $T$ if there are nodes $u, u' \in V_T$ for which:

- $u = v_{R_{T'}(i)}$ for a certain $i \geq d$, where $d$ is the depth of the last node in $T$;
- $u'$ is the left-hand sibling of $u$, if there is any such node;
- $\lambda_T(u) = \lambda_T(u')$, i.e. $u$ and $u'$ have the same label;
- $img_{T'\to G}(u) \supseteq img_{T[R_{T'}(i)-1]\to G}(u')$.

Here, $T' = T[n-1]$, the tree obtained from $T$ by removing the last node. The correctness for this test follows from similar arguments as the previous test, while also borrowing ideas from unavoidable redundancy pruning.

Note that as a consequence of this pruning, certain trees are not entered in the trie data structure. We perform the final maximality checking still on this trie data structure, which means that we need to make sure that we search over missing patterns in the post-processing step.

## 6 Label Constraints

In several of the datasets we will be using in the experiments we can distinguish two types of nodes. On the one hand we have nodes of a general label type, such as "Page", which represent entities in the data; on the other hand, we have nodes that indicate properties of nodes, such as being in the "history" category. The following is a small example of such a database.

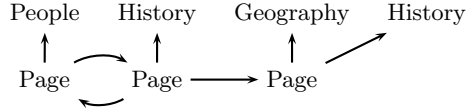

In practice many patterns can be found in such data which contain the "Page" label. From a user's perspective such patterns may not be very useful. Therefore, we propose to impose an additional constraint on patterns.

**Definition 9.** *Let $\mathcal{P}$ be the set of property-labels in a graph. A tree $T$ satisfies the* property label constraint *iff every internal node has a child labeled with a label in $\mathcal{P}$.*

The obvious way to process this constraint is by means of post-processing. A better approach is based on exploiting *unavoidable* violations of the constraint.

**Definition 10.** *Tree $T$ violates the label constraint unavoidably if it has a node that (1) is not on the right-most path and (2) does not have a child labeled with a label in $\mathcal{P}$.*

The processing of this constraint is similar to that of the *unavoidable reducibility* property (Property 4); any tree which unavoidably violates the property label constraint cannot be extended into a tree which satisfies this constraint. Ordering the labels such that property labels are the right-most children of their parent is beneficial for this pruning.

Note that with this type of pruning, we do not achieve incremental polynomial time enumeration of all patterns satisfying the property label constraint: we may be considering many paths before finding a valid pattern. In theory, our algorithm can be modified to also ensure incremental polynomial enumeration in this case; we omit the details of this rather complex modification here. Experiments will be performed for the simpler algorithm.

# 7 Related work

There is a very large literature on frequent pattern mining, ranging from itemsets [1] to tree and graph data [4, 20, 16]. Most work in the graph mining literature focuses on the subgraph isomorphism setting, extending this from molecules to networks [20, 13]. Although in most cases exponential algorithms are used, it was found that certain types of graphs can be mined more efficiently: graphs of bounded treewidth and bounded degree can be mined in incremental polynomial time [10], while outerplanar graphs can be mined with polynomial delay [11], both under isomorphism and homomorphism. The key observation in these papers is to limit the class of graphs that are mined. In our work we do not limit the class of graphs that can be mined; we only limit the class of patterns to trees. In the tree mining literature, additional relationships between data and patterns have been studied, such as embedded subtrees [4]. These methods exploit the ancestor-descendant relationships between nodes in the data, and hence are not applicable when the data is replaced by an arbitrary directed graph. These algorithms have been extended to directed acyclic graphs [19]. In general finding subgraph isomorphisms of trees into DAGs is already an NP complete problem; our algorithm is the first to allow mining of trees into DAGs in incremental polynomial time.

Related to our work are approaches for discovering frequent *conjunctive queries*. In the framework of conjunctive query mining, a pattern is expressed as a *Datalog* query; the data consists of a relational database. The *support* of a Datalog query has been defined in two alternative ways. One is by means of *object identity (OI) subsumption*, which is related to graph isomorphism. The other is by means of *θ-subsumption*, a concept developed in the inductive logic programming (ILP) community which is similar to graph homomorphism. The most-well known conjunctive query miner that exploits $\theta$-subsumption is WARMR [5]. WARMR however prunes patterns which are reducible immediately, and hence is not capable of finding certain patterns. Most systems use OI subsumption to avoid this problem.

An alternative approach, which takes more a database perspective, is the approach for mining *tree queries* [8, 9]. This approach limits itself to conjunctive queries which can be visualized as trees, and hence is similar to our work. Algorithmically, however, it proceeds in two phases: phase one, in which all unlabeled trees up to a certain size are enumerated (avoiding isomorphic trees), and phase two, in which each such tree is refined by adding labels and applying the Apriori principle. Clearly, in this approach many redundant patterns are enumerated at first which may never lead to frequent core patterns. In our approach we can prove that the amount of such patterns considered is bounded by the numbers of patterns already enumerated; furthermore, we deal with cycles in the data.

A recent study also explored the problem of enumerating trees and graphs of bounded tree width under homomorphism [17]. This study focused on the enumeration process without taking into account constraints, such as a support constraint. Essentially, this study showed that without constraints core trees can be enumerated in polynomial delay, i.e., the amount of time spent between two solutions is polynomial in the size of the input. However, the proof relied on the observation that any core tree can be expanded into another core tree within a polynomial number of steps. In frequent pattern mining we have the additional requirement that patterns need to be frequent; also this other recent study [17] lists as future work whether it can be shown that frequent core trees can be enumerated with polynomial delay.

## 8 Experiments

We perform experiments on the following datasets:

**ILPNet** The ILPNet network represents an authorship network of papers, and was collected as part of a European project.[18] In our ILPNet network we included the following nodes and edges:

- nodes for authors, each labeled with a generic 'author' label;
- nodes for papers, each labeled with a generic 'paper' label;
- edges for links between papers and authors (in both directions);
- for each institute associated to an author, a node representing that institute for that particular author; the institute node is labeled with the institute and linked with an edge to the author node, directed from the author to the institute;
- for each keyword associated to a paper, a node representing that keyword for that particular paper; the keyword node is labeled with the keyword and linked with an edge to the paper node, directed from the paper to the keyword.

The resulting network consists of 2 998 nodes, 7 906 edges and 186 labels.

**WebKB** The WebKB network represents a subset of webpages of a number of American universities.[2] The network was processed similarly. We included the following nodes and edges in our network:

- nodes for webpages, each labeled with a type of webpage (course, project, ...);
- single directional edges for links between webpage nodes;
- for each keyword associated to a webpage, a node representing that keyword for that particular webpage; the keyword node is labeled with the keyword and linked with an edge to the webpage node, directed from the webpage to the keyword.

Our dataset includes all 4 different universities and consists of 81 068 nodes, 160 368 edges and 781 labels.

**Wikipedia Schools** The Wikipedia schools network[3] consists of a selection of Wikipedia pages. The dataset consists of the following nodes and edges:

- nodes for Wikipedia pages, each labeled with a generic page label;
- single directional edges for links between pages;
- for each category associated to a page, a node representing the category for that particular page; the category node is labeled with the category and linked with an edge to the page node, directed from the page to the category.

This dataset contains 11 700 nodes, 109 216 edges and 16 labels.

On these datasets we used the following default parameters for our algorithm. On ILP, we applied a minimum support threshold of 20, a maximum pattern size of 20; on WebKB the maximum pattern size was 5, the minimum support 100; on WP the minimum support constraint was 40; the maximum depth constraint 3. Only on the WP data the property label constraint was applied as well. In all cases, the additional cover constraint was used as path constraint. Figures 3 and 4 provide an overview of our results, in which all parameters are set to their default, except those mentioned otherwise in the Figure.

We wish to compare our homomorphism-based algorithm with an isomorphism-based algorithm for mining networks. Unfortunately, implementations of algorithms such as gSpan [20] and its extensions to networks [2] do not support the discovery of rooted trees in directed graph data. Instead, we decided to extend our homomorphism-based implementation to isomorphism-based pattern mining. We implemented two variations: "i-eval", in which patterns are mined under a homomorphism constraint, but support is also evaluated by means of isomorphism, and "i-mine", in which we disabled the homomorphism-based redundancy tests, and only a support count using an exponential subgraph isomorphism algorithm was used, similar to other graph miners [16].

---

[2] http://alchemy.cs.washington.edu/data/webkb/
[3] http://schools-wikipedia.org/

Our algorithms were implemented in Java.

**Q1: Runtime of Isomorphism vs Homomorphism** Figure 3(a) and 3(b) (without property label constraint) show the runtime of both approaches when increasing the maximum size of patterns; increasing the size of patterns increases the branching factor in isomorphism-based algorithms. The difference between "i-eval" and "i-mine", which only is due to the computation of isomorphism-based support, shows that significantly more computation time is needed to determine isomorphism-based support. The runtime increases further when mining only under isomorphism; this is due to large numbers of candidate patterns being generated that are not frequent, which are all evaluated using an isomorphism test. Figure 3(g) shows that the number of patterns found on ILP using homomorphism is actually higher than those found under isomorphism. The graph suggests that filtering by means of homomorphism may be useful even if one is mining for patterns under isomorphism.

**Q2: Delay between Patterns** We proved that our algorithm is incrementally polynomial. Ideally, however, the amount of time spent between two solutions would not depend on the number of solutions listed till that moment.

Figures 3(c) and (d) show a run of our algorithm for two datasets. Plotted is the number of infrequent or non-core patterns enumerated between each pair of frequent core patterns in the output, as the search progresses. On both datasets we can observe that this number is not very dependent on the number of patterns enumerated; indeed, in many cases, the enumeration process only considers between 5 and 20 patterns that are not a core or that are not frequent between each pair of frequent core patterns. This shows that our algorithm performs better in practice than the theory suggests.

**Q3: Condensed Representations** Note that we found a large number of patterns in some of our previous experiments. Figures 3(e) and (f) show the reduction in patterns if we apply closedness and maximality constraints when varying support. Taking into account the logarithmic scale, we can see that the number of patterns found reduces significantly on the ILP dataset, but not on the WebKB dataset; WP is similar to WebKB (not shown). When varying the size of patterns in Figure 3(g), we observe a similar reduction of the number of patterns on ILP data; in this figure we also see that this effect is larger under homomorphism than under isomorphism.

**Q4: Property Label Constraint** We performed experiments with and without the property label constraint. Only a runtime comparison is shown here; Figure 3(b) shows that the runtime improves by our pruning test.

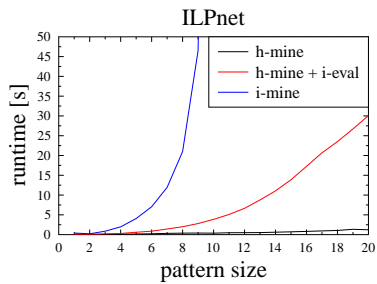
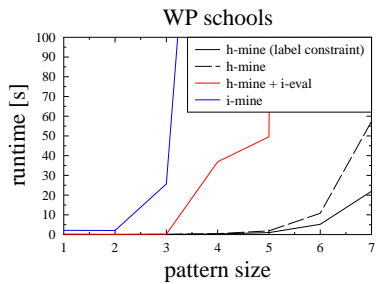

(a) Runtime: Iso- vs Homomorphism    (b) Runtime: Iso- vs Homomorphism

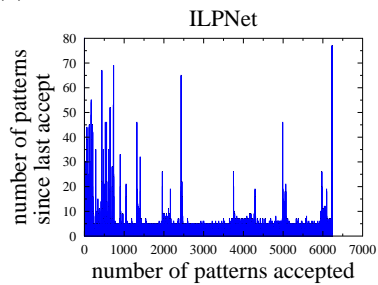
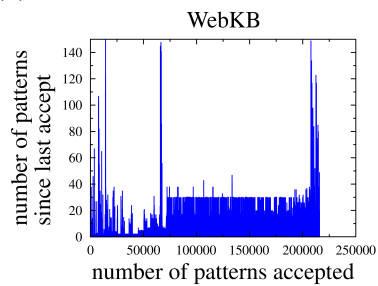

(c) Delay    (d) Delay

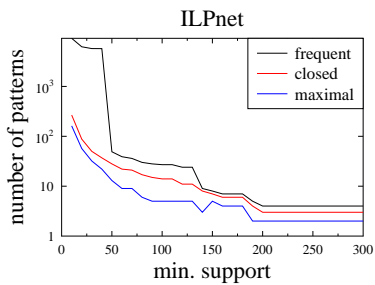
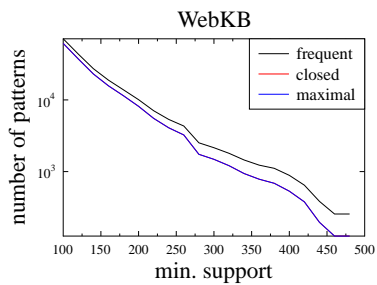

(e) Condensed representations    (f) Condensed representations

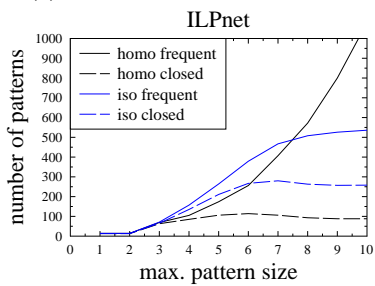
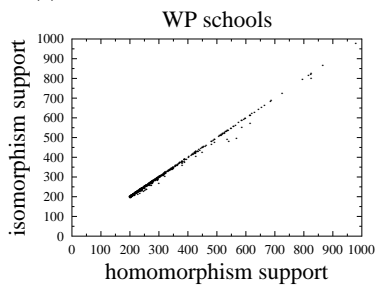

(g) Condensed representations    (h) Comparing supports

Figure 3: Results for the remaining results

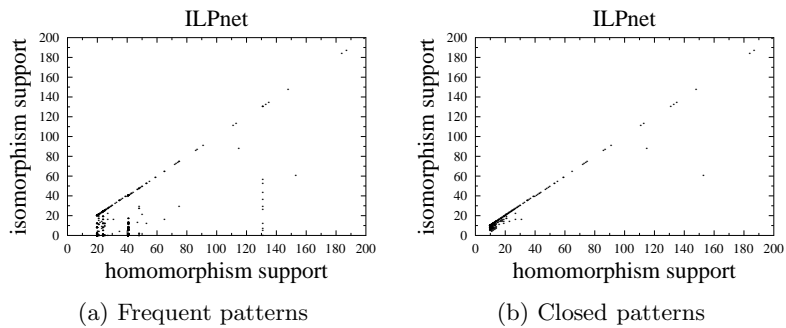

(a) Frequent patterns     (b) Closed patterns

Figure 4: Comparing supports of patterns on the ILP data

**Q5: Differences in Patterns Found**   The experiments in Figures 3(h), 4(a) and 4(b) provide insight in the differences between patterns found in both settings. Each dot represents a pattern found when mining using homomorphism, but also evaluating under isomorphism. The experiments clearly show that both methods provide similar support counts. In ILP we see that there are some sets of patterns that have diverging support values (around homomorphism support 130). These are mainly patterns that are not closed; they are all subtrees of a larger tree that has a similar support both under isomorphism and homomorphism.

Considering closed patterns on ILP data, we still can see some patterns with diverging supports. One such pattern is:

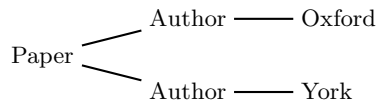

which has an isomorphism-support of 61 but a homomorphism-support of 153. This reflects that papers were written by either an author affiliated to both institutions or by two authors, each affiliated to one of these two institutions.

# 9   Conclusions

In this paper we introduced the problem of mining tree patterns under homomorphism. Experiments showed that its performance is better in practice than the incremental polynomial time delay complexity suggests; typically, the number of infrequent, non-core patterns generated between each pair of frequent, core patterns is small. This shows that our algorithm is also of practical relevance.

Our experiments clearly showed that support evaluation for homomorphism is significantly faster than using isomorphism and hence is beneficial on large databases. Furthermore, in practice the support of patterns found using both types of mappings is typically very similar.

We found that in most cases the number of closed patterns found using homomorphisms is smaller than using isomorphisms. The type of pattern that is not found using our algorithm typically encodes a degree constraint. Where such patterns are undesirable, a possibility is to use our enumeration procedure to filter patterns. In other cases, a possibility for future work is to search for patterns that explicitly encode degree constraints while still using homomorphisms. Arguably, such patterns would be easier to interpret than patterns that encode degree implicitly.